\documentclass[
    ,final            % use final for the camera ready runs
  ]
  {aipproc}

\layoutstyle{6x9}

\newcommand{\ba}{\begin{eqnarray*}}
\newcommand{\ea}{\end{eqnarray*}}
\newcommand{\beq}{\begin{equation}}
\newcommand{\eeq}{\end{equation}}
\newcommand{\bm}[1]{\mbox{\boldmath $#1$}}

\newcommand{\text}{\hbox}

\newcommand{\simordertwo}{\raisebox{-4pt}{$\, \stackrel{\textstyle <}{\sim} \,$}}

\begin{document}

\title{Handedness inside the proton\footnote{Presented at the 15th 
International Spin Physics Symposium (SPIN2002), Brookhaven National 
Laboratory, September 9-14, 2002}}

\author{Dani\"{e}l Boer\\[-3 mm]}{address={Department of Physics and Astronomy, 
Vrije Universiteit Amsterdam\\
De Boelelaan 1081, NL-1081 HV Amsterdam,
The Netherlands\\[-3 mm]}
}

\begin{abstract}
The transversity of quarks inside unpolarized hadrons and its phenomenology
are discussed. Several experimental suggestions are proposed that would
allow further study of this intrinsic handedness.\\[-10 mm]  
\end{abstract}

\maketitle

%%%%%%%%%%%%%%%%%%%%%%%%%%%%%%%%%%%%%%%%%%%%
%% MAINMATTER
%%%%%%%%%%%%%%%%%%%%%%%%%%%%%%%%%%%%%%%%%%%%

\section{Introduction}

As pointed out in Ref.\ \cite{Boer:1999mm} 
there exists an experimental indication 
--a $\cos 2\phi$ azimuthal asymmetry in the Drell-Yan process-- for 
nonzero transversity of quarks inside {\em unpolarized\/} hadrons. The idea 
is that transverse polarization of a noncollinear quark inside an unpolarized 
hadron in principle can have a preferred direction and therefore, does not 
need to average to zero. This preferred direction signals an intrinsic 
handedness. For example expressed in the infinite momentum frame, 
the transverse quark polarization is orthogonal to the directions of 
the proton and the (noncollinear) quark momentum:
\beq
{S_{T}^q \, \propto \, P_{\text{hadron}} \times p_{\text{quark}}}.
\eeq
Clearly, this must be related to {orbital angular momentum}, but how exactly 
is still an open question. 

Whether there is indeed nonzero intrinsic handedness remains to be tested 
and here we will discuss ways of how one would be able to pursue
this issue. Some theoretical aspects of this quark-transversity
distribution function (usually denoted by
$h_1^\perp$) \cite{Boer:1998nt} will be reviewed and its main experimental
signatures will be pointed out. 
In particular, unpolarized and single spin asymmetries will be
discussed for the Drell-Yan (DY) 
process and semi-inclusive DIS. Important in the 
latter case are polarized $\Lambda$ production observables. 
Special emphasis will be put on how to distinguish the various asymmetries
compared to those arising from other mechanisms, like the Sivers effect 
\cite{Sivers:1990cc}.\\[-9 mm] 

\section{``T-odd'' distribution functions}

The quark-transversity function  
$h_1^\perp$ is a function of the lightcone momentum fraction $x$ and 
transverse momentum $\bm{p}_T$ of a quark inside an unpolarized hadron. 
At first sight, this intrinsic handedness function 
appears to violate time reversal invariance, when the incoming hadron is 
treated as a {plane-wave state}. However, already in a simple gluon-exchange 
model calculation \cite{Brodsky:2002cx} such so-called ``T-odd'' 
distribution functions turn out to be nonzero. 
In recent work \cite{Collins:2002kn,Ji:2002aa,Belitsky:2002sm} 
it has been demonstrated that the proper gauge invariant definition of
transverse momentum dependent functions does indeed allow for such 
seemingly time reversal symmetry violating functions\footnote{The name
``T-odd'' is thus 
a misnomer, since it seems to suggest a violation of time reversal 
invariance, which turns out not to be the case. 
Other names, like naive or artificial T-odd, have been
suggested.}. Therefore, here 
we will simply assume that there exists no symmetry argument that forces the
intrinsic handedness to be absent. 

A large part of the $h_1^\perp$ 
phenomenology was already presented in Refs.\ \cite{Boer:1998nt,Boer:1999mm}.
The function  
$h_1^\perp$ enters the 
asymmetries discussed below without suppression by inverse powers of the hard
scale in the process (but the operator associated to the function is not 
twist-2 in the OPE sense). 
The other unsuppressed transverse momentum dependent ``T-odd'' function is the
Sivers effect function, denoted by $f_{1T}^\perp$. It parameterizes the
probability of finding an unpolarized quark (with $x$ and $\bm{p}_T$) inside 
a transversely polarized hadron.  

\section{Unpolarized Drell-Yan}

As mentioned, there exists data that is {compatible} with nonzero 
${h_1^\perp}$. 
A large $\cos 2\phi$ angular dependence in the unpolarized DY process 
$\pi^- N \rightarrow \mu^+ \mu^- X$ was observed, for deuterium and tungsten
and with $\pi$-beam energies ranging between 140 and 286 GeV
\cite{Falciano:1986wk,Guanziroli:1988rp,Conway:1989fs}. 
Conventionally, the differential cross section is written as
\beq
\frac{1}{\sigma}\frac{d\sigma}{d\Omega} \propto 
\left( 1+ \lambda \cos^2\theta + \mu \sin^2\theta \, 
\cos\phi + \frac{{\nu}}{2} \sin^2 \theta \, {\cos 2\phi} \right),
\eeq
where $\phi$ is the angle between the lepton and hadron scattering planes in
the lepton center of mass frame (see Fig.\ 3 of Ref.\ \cite{Boer:1999mm}).
The perturbative QCD (pQCD) prediction for very small transverse momentum
($Q_T$) of the muon pair is $\lambda \approx 1, \mu \approx
0, \nu \approx 0$. More generally, i.e.\ also for larger $Q_T$ values, 
one expects the Lam-Tung relation $1-\lambda -2 \nu = 0$ to hold (at order
$\alpha_s$). However, the data (with  
invariant mass $Q$ of the lepton pair in the range $Q \sim 4 - 12$ GeV) 
is incompatible with this pQCD relation (and with its ${\cal O}(\alpha_s^2)$ 
modification as well \cite{Brandenburg:1993cj}). 
Several explanations have been
put forward in the literature, but these will not be reviewed here.
 
In Ref.\ \cite{Boer:1999mm} 
we have observed that within the framework of transverse momentum dependent 
distribution functions, 
the $\cos 2\phi$ asymmetry can only be accounted for by the
function $h_1^\perp$ or else will be $1/Q^2$ suppressed. 
We obtained $\nu \propto h_1^{\perp\, \pi} \, h_1^{\perp\, N}$ and this
expression was used to fit the function $h_1^{\perp}$ from the data.
This approach 
has several aspects in common with earlier work by Brandenburg, 
Nachtmann and Mirkes \cite{Brandenburg:1993cj}, where the large values of
$\nu$ were generated from
a nonperturbative, nonfactorizing mechanism that correlates the transverse
momenta and spins of the quark and anti-quark that annihilate into the virtual
photon (or $Z$, but not $W$, boson). But the description of $\nu$ as 
a product of two $h_1^\perp$ functions implies that these correlations are 
not necessarily factorization breaking and this type of effects will then 
not be specific to 
hadron-hadron scattering. We also note that since the function 
$h_1^{\perp}$ is a quark helicity-flip matrix element, it offers a natural
explanation for $\mu \approx 0$. 

\section{Polarized Drell-Yan}

Instead of colliding two unpolarized hadrons, 
one can also use a polarized hadron to become sensitive to the polarization
of quarks inside an unpolarized hadron. In principle, this 
provides a new way to 
measure the transversity distribution function $h_1$. 
In this case the transverse hadron spin (${\bm S}_T$)
dependent differential cross section may be parameterized by 
(choosing $\mu=0$ and $\lambda=1$)
\beq
\frac{d\sigma(p\, p^\uparrow \to \ell \, \bar \ell \, X)}{d\Omega\; 
d\phi_{S}} \propto 
1+ \cos^2\theta 
+ \sin^2 \theta \left[ \frac{{\nu}}{2} \; \cos 2\phi - {\rho} \; 
|\bm S_{T}^{}|\;
{\sin(\phi+\phi_{S})} \right] + \ldots, 
\eeq
where $\phi_S$ is the angle of the transverse spin compared to the lepton
plane, cf.\ Ref.\
\cite{Boer:1999mm}.

\begin{figure}[htb]
\includegraphics[width = 7 cm]{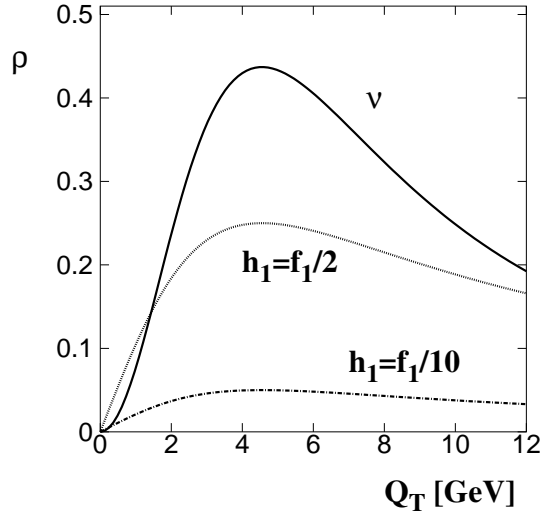}
\vspace{-12pt}
\caption{Results for $\rho$ using Eq.\ (\ref{approx}), 
for $h_1$ equal to $f_1/2$ and $f_1/10$.}
\label{fit2}
\end{figure}

The analyzing power $\rho$ is (within this framework) 
proportional to the product 
$h_1^\perp \, h_1 $ \cite{Boer:1998nt,Boer:1999mm}. 
Hence, the measurement of $\langle\cos 2\phi \rangle$ (e.g.\ at {\small 
RHIC} in 
$p \, p \rightarrow \mu^+ \mu^- X$ or at Fermilab in $p \, \bar p 
\rightarrow \mu^+ \mu^- X$) combined with a measurement of the 
single spin azimuthal asymmetry $\langle {\sin(\phi+ \phi_{S})} \rangle$ 
(also possible at {\small RHIC})
could provide information on $h_1$. In other words, 
a nonzero function $h_1^\perp$ will imply a relation between $\nu$ and $\rho$,
which in case of one (dominant) flavor (usually called $u$-quark 
dominance) and Gaussian transverse momentum dependences, 
is approximately given by 
\beq
\rho \approx
\frac{1}{2} \, 
\sqrt{\frac{{\nu} }{{\nu_{\mbox{${\rm max}$}}}}}\, \frac{h_1}{f_1},
\label{approx}
\eeq
where $\nu_{\mbox{${\rm max}$}}$ is the maximum value attained by $\nu (Q_T)$.
This relation depends on the magnitude of $h_1$ compared to $f_1$ and 
since $h_1$ is not known experimentally, in Fig.\
\ref{fit2} we display two options for $\rho$, using the function 
$\nu$ which was fitted from the 194 GeV data \cite{Boer:1999mm} (which has 
$\langle Q_T \rangle \simordertwo 3$ GeV) and is extrapolated to larger 
$Q_T$ values 
(the theoretically expected turn-over of $\nu$ has not yet been seen
in experiments).

We note that the Sivers function $f_{1T}^\perp$ will generate a 
different angular single spin asymmetry, namely proportional to 
$( 1+ \cos^2\theta )\;|\bm S_{T}^{}|\;
{\sin(\phi-\phi_{S})}\; {f_{1T}^{\perp}} \; f_1$.

\section{Hadron production single spin asymmetries}

Large single transverse spin asymmetries have been observed 
in the process $p \, p^{\uparrow} \rightarrow \pi \, X$ \cite{Adams:1991cs}.
It has been suggested that these asymmetries can arise 
from various ``T-odd'' functions with 
transverse momentum dependence. There are three options:
\[
{h_1^\perp} \otimes h_1 \otimes D_1; \quad
{f_{1T}^\perp} \otimes f_1  \otimes D_1; \quad
h_1  \otimes f_1 \otimes H_1^\perp.
\]
The first two options are similar to those described in the previous
section, now accompanied
by the unpolarized fragmentation function $D_1$. The third option contains 
the Collins effect function $H_1^\perp$ \cite{Collins:1993kk}, 
which is 
the fragmentation function analogue of $h_1^\perp$, but in principle is
unrelated in magnitude. The last two options were investigated in 
\cite{Anselmino:1995tv,Anselmino:1999pw}.
 
We note that the first two options also occur in jet 
production asymmetries:
{$p + p^{\uparrow} \rightarrow \text{jet} + X$} (with only neutral current
contributions for the first option). However, as Koike has pointed out, 
the first option is a double transverse
spin asymmetry on the parton level and is thus expected to be small, like the
example of Ref.\ \cite{Kanazawa:2000kp} or the 
double transverse spin asymmetry in DY. It is therefore more likely to
obtain information on $h_1^\perp$ from hadron production asymmetries in 
semi-inclusive DIS (SIDIS). 

\section{Semi-inclusive DIS}

First some comments on unpolarized asymmetries in SIDIS. The 
$\langle {\cos 2\phi} \rangle$ in SIDIS at values of $Q^2$ similar to
those of the unpolarized DY data, has been measured by the EMC
collaboration \cite{Aubert:1983cz,Arneodo:1987cf}. 
No significant asymmetry was observed due to the
large errors. But in the present picture of ``T-odd'' functions 
the $\langle {\cos 2\phi} \rangle$ in {SIDIS}
would be {$\propto h_1^\perp H_1^\perp$}, which thus can be quite different in
magnitude. A consistent picture should emerge by also comparing to 
$\langle {\cos 2\phi} \rangle \propto H_1^\perp H_1^\perp$ in 
$e^+ e^-$ annihilation 
(e.g.\ doable at {\small BELLE}). However, 
one should keep in mind that
there is another source of a $\cos 2\phi $ asymmetry, namely one  
that stems from double 
gluon radiation in the hard scattering subprocess. Fortunately, this forms a 
calculable background which only 
dominates in the large $Q_T$ region (close to $Q$). 
 
Another test would be to look at $\langle {\cos 2\phi} \rangle$ for a
jet instead of a hadron: {$e \, p \to e' \, \text{jet} \, X$}.  The
contribution from $h_1^\perp$ will then be absent.
  
Apart from these unpolarized asymmetries, one can also consider polarized
hadron production asymmetries in SIDIS, most notably 
polarized $\Lambda$ production. In that case the intrinsic handedness can 
lead to the following asymmetries:
\begin{itemize}
\item {$\sin(\phi^e_\Lambda 
+ \phi^e_{S_T^\Lambda})$} and $\sin(3\phi^e_\Lambda - 
\phi^e_{S_T^\Lambda})$ in 
{$e\, p \to e' \, \Lambda^{\uparrow} \, X$} \hspace{2 mm} 
(transverse $\Lambda$
polarization) 
\item {$\sin(2\phi^e_\Lambda)$} in {$e\, p \to e'\,
\vec{\Lambda} \, X$} \hspace{2 mm} (longitudinal $\Lambda$
polarization) 
\end{itemize}
These particular 
angular dependences should be {absent} for {charged 
current exchange processes}, 
like $\nu \, p \to e \, \Lambda^{\uparrow} \, X$ or $\nu \, p \to e \, 
\vec{\Lambda} \, X$ \cite{Boer:1999uu}.

These asymmetries are distinguishable from other mechanisms via the 
$y$ and $\phi^e$ dependences. For instance, the first asymmetry for 
transversely
polarized $\Lambda$ production, can be distinguished from the asymmetry due to
the so-called {polarizing fragmentation functions}
\cite{Mulders:1996dh,Anselmino:2000vs} (also called the Sivers
fragmentation function, although it is in principle unrelated in magnitude
to the Sivers distribution function).
Moreover, the asymmetries should vanish after integration over $Q_T$, 
leaving only possibly a {$\sin(\phi^e_{S_T^\Lambda})$} asymmetry (which is 
a {twist-3}, and hence suppressed, asymmetry) \cite{Kanazawa:2000cx,Koike-02}.
 
\section{Conclusions}

The chiral-odd, ``T-odd'' 
distribution function $h_1^\perp$ offers an
explanation for the large unpolarized $\cos 2 \phi$ 
asymmetry in the $\pi^- N \rightarrow \mu^+
\mu^- X$ data. Nonzero $h_1^\perp$ would 
relate unpolarized and polarized observables in a distinct way and thus 
in principle offers {a new way 
to access $h_1$} in $p \, p^\uparrow \rightarrow \mu^+ \mu^- X$. 

There are {several ways of differentiating $h_1^\perp$ dependent 
asymmetries from those due to 
other mechanisms} and the suggestion of nonzero $h_1^\perp$ can be explored 
using a host of 
existing (Fermilab, {\small BELLE}) and near-future data 
({\small RHIC, COMPASS, HERMES}). 

%%%%%%%%%%%%%%%%%%%%%%%%%%%%%%%%%%%%%%%%%%%%%%%%
%% BACKMATTER
%%%%%%%%%%%%%%%%%%%%%%%%%%%%%%%%%%%%%%%%%%%%%%%%

\begin{theacknowledgments}
I thank Arnd Brandenburg, Stan Brodsky, Dae Sung Hwang, 
Yuji Koike and Piet \mbox{Mulders} for fruitful discussions on this topic. 
The research of D.B.~has been made possible by 
financial support from the Royal Netherlands Academy of Arts and Sciences.

\end{theacknowledgments}

%% If the bibliography is produced without BibTeX comment out the

\bibliographystyle{aipproc}   % if natbib is available

\bibliography{SPIN2002chepph}

\IfFileExists{\jobname.bbl}{}
 {\typeout{}
  \typeout{******************************************}
  \typeout{** Please run "bibtex \jobname" to optain}
  \typeout{** the bibliography and then re-run LaTeX}
  \typeout{** twice to fix the references!}
  \typeout{******************************************}
  \typeout{}
 }

\end{document}